\documentclass[
aip,
 amsmath,amssymb,
preprint,%
 reprint,%
]{revtex4-1}

\usepackage{graphicx}% Include figure files
\usepackage{dcolumn}% Align table columns on decimal point
\usepackage{bm}% bold math
\usepackage[utf8]{inputenc}
\usepackage[T1]{fontenc}
\usepackage{mathptmx}
\usepackage{etoolbox}

\makeatletter
\def\@email#1#2{%
 \endgroup
 \patchcmd{\titleblock@produce}
  {\frontmatter@RRAPformat}
  {\frontmatter@RRAPformat{\produce@RRAP{*#1\href{mailto:#2}{#2}}}\frontmatter@RRAPformat}
  {}{}
}%
\makeatother
\begin{document}

\preprint{AIP/123-QED}

\title{Development of microwave surface elastoresistivity measurement technique under tunable strain
}
% Force line breaks with \\
\author{Suguru Hosoi}
\thanks{Authors to whom correspondence should be addressed:shosoi@lanl.gov}

 \affiliation{Graduate School of Engineering Science, Osaka University, Toyonaka, Osaka, 560-8531, Japan.}%\\This line break forced with \textbackslash\textbackslash

 \affiliation{Los Alamos National Laboratory, Los Alamos, New Mexico, 87545, USA}%Lines break automatically or can be forced with \\

\author{Kohei Matsuura}%
\thanks{matsuura@qipe.t.u-tokyo.ac.jp}

% \email{matsuura@qipe.t.u-tokyo.ac.jp}
  \affiliation{ 
Department of Applied Physics, Graduate School of Engineering, The University of Tokyo, Bunkyo-ku, Tokyo, 113-8656, Japan.
%\\This line break forced with \textbackslash\textbackslash
}%

\author{Masaaki Shimozawa}
 \affiliation{Graduate School of Engineering Science, Osaka University, Toyonaka, Osaka, 560-8531, Japan.%\\This line break forced with \textbackslash\textbackslash
}

\author{Koichi Izawa}
 \affiliation{Graduate School of Engineering Science, Osaka University, Toyonaka, Osaka, 560-8531, Japan.%\\This line break forced with \textbackslash\textbackslash
}

\author{Shigeru Kasahara}
 \affiliation{Research Institute for Interdisciplinary Science, Okayama University, Okayama, 700-8530, Japan%\\This line break forced with \textbackslash\textbackslash
}

\author{Takasada Shibauchi}
 \affiliation{Department of Advanced Materials Science, The University of Tokyo, Kashiwa, Chiba 277-8561, Japan%\\This line break forced with \textbackslash\textbackslash
}

\date{\today}% It is always \today, today,

\begin{abstract}
By integrating a dielectric microwave resonator with a piezoelectric-based strain device, we develop an in situ strain-tunable microwave spectroscopy technique that enables contactless measurements of superconducting properties under strain. In the slightly overdoped iron-based superconductor BaFe$_2$(As$_{1-x}$P$_x$)$_2$, we successfully observe a systematic strain dependence of the superconducting transition, manifested as changes in the quality factor and resonance frequency shifts. Both compressive and tensile anisotropic lattice distortions along the [110]${_{\rm T}}$ direction suppress superconductivity, consistent with standard transport measurements, highlighting the pivotal role of nematic fluctuations in the superconducting mechanism. Our strain-tunable cavity therefore serves as a powerful, contactless probe of fundamental superconducting material properties under strain and may also potentially facilitate the design of hybrid quantum systems with strain-controlled quantum degrees of freedom.
 % We demonstrate in situ strain-tunable microwave spectroscopy measurements by integrating a dielectric microwave resonator with a piezoelectric-based strain device. Our strain cavity allowed microwave impedance measurements under strain in the slightly overdoped iron-based superconductor BaFe$_2$(As$_{1-x}$P$_{x}$)$_2$, where we successfully observed the systematic strain dependence of superconducting transition through changes in quality factor and resonance frequency shifts. Both compressive and tensile anisotropic lattice distortions along the [110]$_{\rm T}$ direction suppressed superconductivity, highlighting the pivotal role of nematic fluctuations in the superconducting mechanism. Mechanically strain-tunable resonators can thus serve as powerful probes of fundamental material properties and potentially facilitate the design of hybrid quantum systems with strain-controlled quantum degrees of freedom.
\end{abstract}

\maketitle

%%%%%%  Introduction %%%%%%%%%
 % Elastic strain has become widely utilized as a symmetry-breaking force to tune quantum materials. 
 Strain engineering has proven effective in controlling various material functionalities across diverse fields of physics, including the tuning of magnetization\cite{SOtaNatEle2018}, spin transport\cite{PhysRevApplied.18.024005}, valley degrees of freedom\cite{PhysRevResearch.6.033096}, and topological properties\cite{PhysRevLett.123.036806}.
 % One of the unique capabilities of elastic strain is its ability to lower the point group symmetry of the lattices through anisotropic stress. In iron-based superconductors with a tetragonal structure, applying small perturbative anisotropic strains that induce orthorhombic distortions \textemdash achievable by simply attaching the sample on the surface of a piezo actuator \textemdash has been demonstrated to probe the critical enhancement of electronic nematic susceptibility in various iron-based superconductors\cite{doi:10.1126/science.1221713,doi:10.1126/science.aab0103,doi:10.1073/pnas.1605806113,doi:10.1073/pnas.2110501119}. 
 In particular, the development of piezoelectric tunable strain devices offers the potential to apply large strain beyond $1 \%$\cite{annurev:/content/journals/10.1146/annurev-conmatphys-040521-041041}, enabling the exploration of novel quantum phenomena under systematic strain control.
  % Strain engineering has proven effective in controlling various material functionalities across diverse fields of physics, including the tuning of magnetization\cite{SOtaNatEle2018}, spin transport\cite{PhysRevApplied.18.024005}, valley degrees of freedom\cite{PhysRevResearch.6.033096}, and topological properties\cite{PhysRevLett.123.036806}.
 Notable examples include the sharp peak of superconducting transition temperatures observed in Sr$_2$RuO$_4$\cite{doi:10.1126/science.aaf9398} and the uniaxial-stress induced charge density wave in high-$T_{\rm c}$ cuprate superconductors\cite{doi:10.1126/science.aat4708}. 

 % Successful strain-tuning of electronic states highlights the promising role of strain engineering. Indeed, it has been demonstrated across various fields of physics that strain can serve as a powerful tool to control material functionalities, including strain-tuning of magnetization\cite{SOtaNatEle2018}, spin transport\cite{PhysRevApplied.18.024005}, valley degrees of freedom\cite{PhysRevResearch.6.033096}, and topological properties\cite{PhysRevLett.123.036806}. 
 
These advances have significantly broadened the scope of strain-based research, driving the need for novel probes capable of elucidating the electronic properties under strain. To this end, piezoelectric devices have been combined with various experimental techniques, including electric transport\cite{PhysRevLett.120.076602}, specific heat\cite{10.1063/5.0021919}, elastocaloric effects\cite{10.1063/1.5099924}, spectroscopy\cite{PhysRevB.99.035118,PhysRevB.108.144501}, and scattering\cite{10.1063/5.0025307}.
  However, non-contact, high-sensitivity probes capable of detecting low energy quasiparticle excitations\textemdash essential for understanding superconducting properties such as pairing symmetry, gap structures, and coexisting order within the superconducting phase \textemdash remain limited. While specific heat measurements serve as one possible approach, achieving sufficient resolution under strain is challenging due to the inevitable strong thermal coupling to the strain device. In contrast, the microwave spectroscopy technique employed here is inherently contactless and highly sensitive to various collective excitations within materials, including spin excitations\cite{PhysRevLett.113.083603,PhysRevLett.113.156401} and superconducting quasiparticles\cite{PhysRevLett.88.047005,PhysRevLett.102.207001}. Integrating strain control techniques with microwave spectroscopy measurements thus provides a promising route to explore low-energy quasiparticle dynamics in strain-tuned quantum materials.

  % Microwave cavities are among the most effective and versatile tools for technological uses in quantum processing as well as fundamental investigations in condensed matter physics. They have, for instance, become indispensable platforms for hosting and operating superconducting qubits\cite{devoretreview}.
  % In fundamental physics research, microwave spectroscopy is highly sensitive to various collective excitations within materials, including spin excitations\cite{PhysRevLett.113.083603,PhysRevLett.113.156401} and superconducting quasiparticles\cite{PhysRevLett.88.047005,PhysRevLett.102.207001}. In particular, the microwave surface impedance of superconductors should be a valuable probe for elucidating the superconducting properties since DC resistivity becomes zero in the superconducting state.

 % Integrating mechanical control, such as pressure, is a promising method to enhance the performance of microwave resonators. Recently, inducing strain into superconducting quantum circuits on the platform successfully improved the coherence time of qubits\cite{doi:10.1126/science.1226487,LisenfeldNatCom,BilmesnpjQI}. However, in previous methods, strain was applied to both superconducting qubits and microwave resonator circuits, making it difficult to precisely control the microwave properties or pinpoint the essential role of strain on microwave resonance. 
 
  In this study, we developed a strain-tunable microwave cavity using high-dielectric rutile, enabling selective strain application to the target by isolating the microwave transmission line from the strain device. To demonstrate the capability of this device, we performed strain-dependent microwave spectroscopy measurements of the iron-based superconductor BaFe$_2($As$_{1-x}$P$_x)$$_2$. Iron-based superconductors are ideal targets for this purpose owing to their particular sensitivity to uniaxial stress applied along the specific crystal directions that couple to the direction of their electronic nematic order\cite{doi:10.1126/science.aab0103}. By using our  strain-tunable microwave cavity, the superconducting transition was successfully observed through the temperature dependence of the quality factor and resonance frequency shifts. Moreover, strain-dependent microwave spectroscopy measurements revealed the suppression of superconductivity under both compressive and tensile strain. These results support the scenario that nematic fluctuations, which are suppressed by the applied anisotropic strain, play a crucial role in enhancing iron-based superconductivity\cite{PaulNatPhys}. Furthermore, we successfully measured microwave surface elastoresistivity, defined as the strain-induced change in microwave surface resistivity. The observed microwave elastoresistivity exhibits a strain dependence similar to that of standard elastoresistivity, demonstrating its potential as an alternative tool to study the anisotropic properties of materials. From a technical perspective, our compact strain-tunable cavity design, which fits within a 26 mm diameter bore system, is compatible with future operations under standard high static magnetic fields, expanding its potential for advanced applications.

 \begin{figure}
  \includegraphics[width=0.48\textwidth]{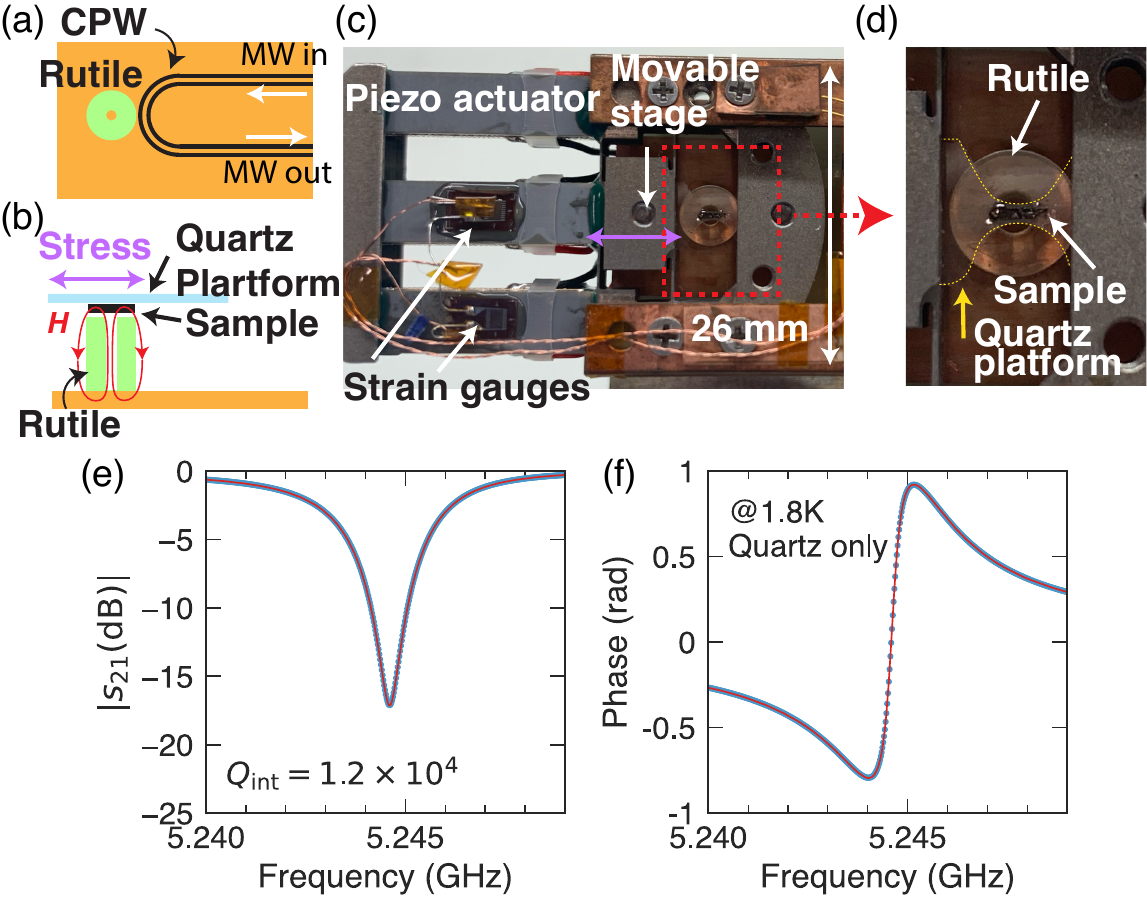}% Here is how to import EPS art
  \caption{\label{fig:ExpSetup} (a) and (b) Schematic top (a) and side (b) views of the rutile cavity with a 2D coplanar waveguide (CPW) platform. The hollow cylindrical rutile crystal is placed near the microwave (MW) transmission line, and the scattering parameter $s_{\rm 21}$ is measured to detect the resonance of rutile. The sample, located just above the rutile, is glued to the underside of the quartz platform to apply uniaxial stress. (c) Photograph of the strain-tunable microwave cavity. The device is compact enough to fit within a 26 mm bore. By applying voltage to piezo actuators, in situ strain tuning is achieved. To monitor the stage displacement, resistive strain gauge are attached to both inner and outer piezo actuators. (d) Enlarged view of the sample mounting area. (e) and (f) $s_{21}$ of the rutile resonance at 1.8 K, without a sample: the magnitude (e) and phase (f). $Q_{\rm int}$ of our cavity reaches the order of $10^4$.}
\end{figure}

 Figures \ref{fig:ExpSetup}(a)(b) depict the schematic diagrams of our strain-tunable microwave cavity. A high-dielectric rutile crystal is placed near the microwave transmission line, enabling coupling to the coplanar waveguide. Rutile possesses excellent microwave properties, including a very large relative dielectric constant and extremely low loss tangent, making it ideal for constructing high-$Q$ resonators when enclosed. The $Q$ factor of the dielectric rutile resonator reaches $Q\sim10^6$, well beyond that of the copper cavity\cite{10.1063/1.2167127}. Another advantage of using rutile is its ability to strongly confine microwave fields within the small space, making it compatible with the strain-tuning device, as shown in Figs. \ref{fig:ExpSetup}(c)(d). This strain device is driven by three piezoelectric actuators, and both compressive and tensile strain can be applied in situ in the cryostat\cite{10.1063/1.4881611}. The samples are glued to the quartz platform, allowing the application of strain without affecting the main microwave components, including the rutile and the microwave transmission line. This design, in principle, offers the advantage of isolating and pinpointing the essential strain response in the sample's microwave properties.

 The strain cell is made of titanium, selected to compensate for the thermal expansion of the piezo-actuators\cite{10.1063/1.4881611}. However, this design is not ideal for constructing a microwave cavity, as the titanium cell partially encloses the rutile, leading to a reduced quality factor, due to both titanium's low conductivity and incomplete enclosure. Despite these limitations, a clear rutile resonance is observed in the transmission spectrum $s_{21}$, as shown in Figs. \ref{fig:ExpSetup}(e)(f). The internal quality factor approaches $Q_{\rm int} \sim 10^4$, as evaluated using the noise-tolerant analysis method\cite{10.1063/1.4907935} (see also supplementary material). This resolution is sufficient to detect changes in the microwave surface resistance, including those associated with the superconducting transition.

\begin{figure}
  \includegraphics[width=0.48\textwidth]{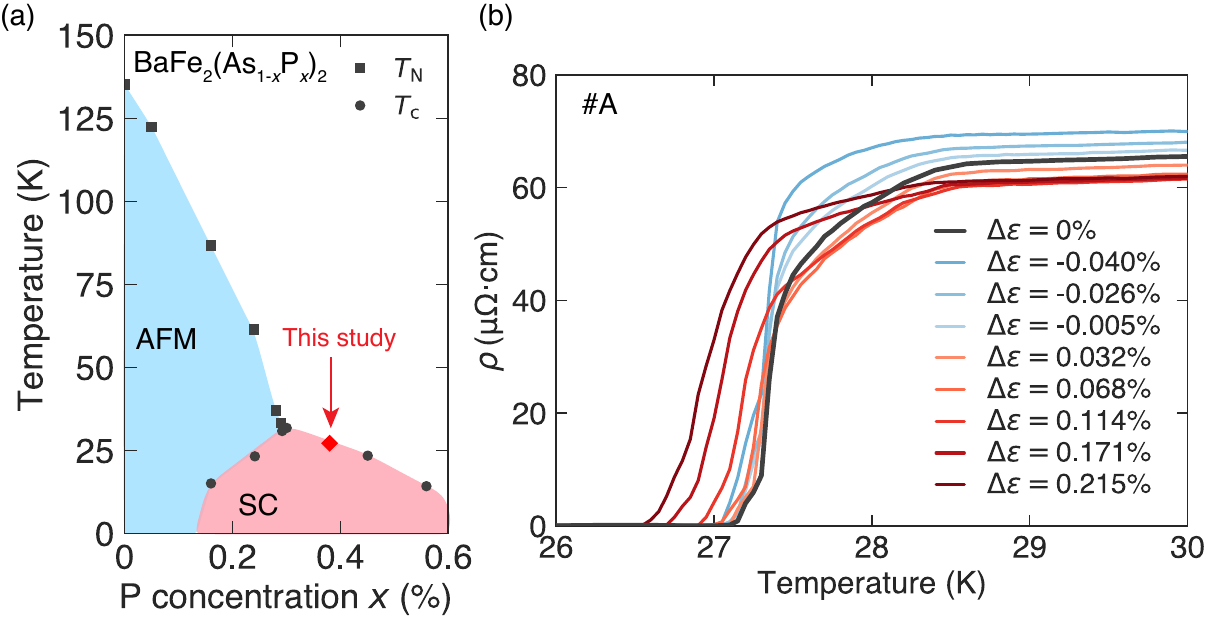}% Here is how to import EPS art
  \caption{\label{fig:BaFe2As2} (a) Phase diagram of BaFe$_2$(As$_{1-x}$P$_x$)$_2$ from Ref. [\onlinecite{doi:10.7566/JPSJ.86.083706}]. The red arrow indicates the slightly overdoped composition at $x \sim 0.38$ used in this study. (b) Strain-dependent resistivity near the superconducting transition for sample $\#$A.}
\end{figure}

 To demonstrate the functionality of our developed strain-tunable cavity, we investigated the strain-dependent superconductivity of the prototypical iron-based superconductor BaFe$_2$As$_2$ system. Figure \ref{fig:BaFe2As2}(a) depicts the phase diagram of BaFe$_2$(As$_{1-x}$P$_x$)$_2$, as an isovalently substituted system with P replacing As sites\cite{doi:10.7566/JPSJ.86.083706}. The parent material BaFe$_2$As$_2$ exhibits antiferromagnetic (AFM) order, accompanied by a tetragonal-to-orthorhombic structural transition (nematic order). Chemical substitution successfully induces superconductivity followed by the suppression of nematicity. Nematicity has been extensively discussed as a potential driving force for superconductivity in iron-based materials\cite{FernandesNatPhys}. Indeed, strongly enhanced nematic fluctuations emerge near the optimal doping across a wide range of iron-based superconductors\cite{doi:10.1126/science.1221713,doi:10.1126/science.aab0103,doi:10.1073/pnas.1605806113,doi:10.1073/pnas.2110501119}. Consistently, superconductivity in optimally-doped Ba(Fe$_{1-x}$Co$_x$)$_2$As$_2$ can be drastically suppressed by the anisotropic strain, which artificially induces orthorhombicity and suppresses nematic fluctuations\cite{PaulNatPhys}. Thus, superconducting states governed by nematic fluctuations are promising candidates for exhibiting high strain tunability. In particular, BaFe$_2$(As$_{1-x}$P$_x$)$_2$ serves as an ideal material for evaluating the effectiveness of our strain-tunable cavity in investigating the effect of strain on superconductivity via microwave spectroscopy.

  To investigate the intrinsic superconducting properties without the influence of magnetism, we focused on the slightly overdoped composition at $x = 0.38$, as determined from $T_{\rm c}$ based on the phase diagram\cite{doi:10.7566/JPSJ.86.083706}(Fig. \ref{fig:BaFe2As2}(a)). To evaluate the strain effects on the superconducting transition, we first performed standard four-probe resistivity measurements, following the same procedure as in the strain-tunable cavity experiments, where the samples were glued onto the quartz platform. A resistive strain gauge was attached on the backside of the platform to quantify the induced strain along [110]$_{\rm T}$. Figure \ref{fig:BaFe2As2}(b) shows strain-dependent resistivity around $T_{\rm c}$, demonstrating the suppression of superconductivity both compressive and tensile strain along $[110]_{\rm T}$. This result is consistent with the Co-doped system\cite{PaulNatPhys}, suggesting that our slightly overdoped samples remain strongly influenced by nematic fluctuations and realize the highly strain-tunable superconducting state. 

  % Next, we performed microwave resonance measurements under strain. 
  We now introduce our central experimental development: a microwave resonance technique capable of probing superconducting properties under strain.
  Two key factors in microwave resonance properties are the quality factor $Q_{\rm s}$ and its resonance frequency $f_Q$. From these values,  microwave surface impedance $Z_{\rm s} = R_{\rm s} + i X_{\rm s}$ can be determined. The real part, the surface resistance $R_{\rm s}$, is related to the microwave loss $1/Q_{\rm s}$ of the sample through the expression $R_{\rm s} = G/Q_{\rm s}$, where $G$ is the geometric factor. This surface resistance $R_s$ qualitatively reflects DC resistivity, and in the limit of $\omega \tau \ll 1$ (where $\omega = 2 \pi f_{Q}$ and $\tau$ is the scattering time), the equation $R_{\rm s} = \sqrt{\mu_0 \omega \rho/2}$ becomes valid. Here, $\mu_0$ is the permeability in vacuum. In this limit, the imaginary part of surface impedance, the surface reactance $X_{\rm s}$, yields the same result: $R_{\rm s} = X_{\rm s}$. In superconducting states, the situation changes drastically. While $R_s$ rapidly decreases, the surface reactance $X_{\rm s}$ can directly probe the magnetic penetration depth $\lambda$ of superconductors, the ratio of effective quasiparticle mass $m^{*}$ to superfluid density $n$ as $\lambda^2 = \frac{m^*}{\mu_0 n e^2}$.  To extract $X_{\rm s}$, temperature dependence of resonance frequency is required:$\Delta X_{\rm s} (T) = - 2 G \frac{\Delta  f_Q (T)}{f_Q (T_{\rm min})}$. Here, $\Delta X_{\rm s} (T)$ and $\Delta  f_Q (T)$ describes changes in $ X_{\rm s}$ and $f_{Q}$ relative to the minimum measurable temperature $T_{\rm min}$, respectively.

\begin{figure}
    \includegraphics[width=0.48\textwidth]{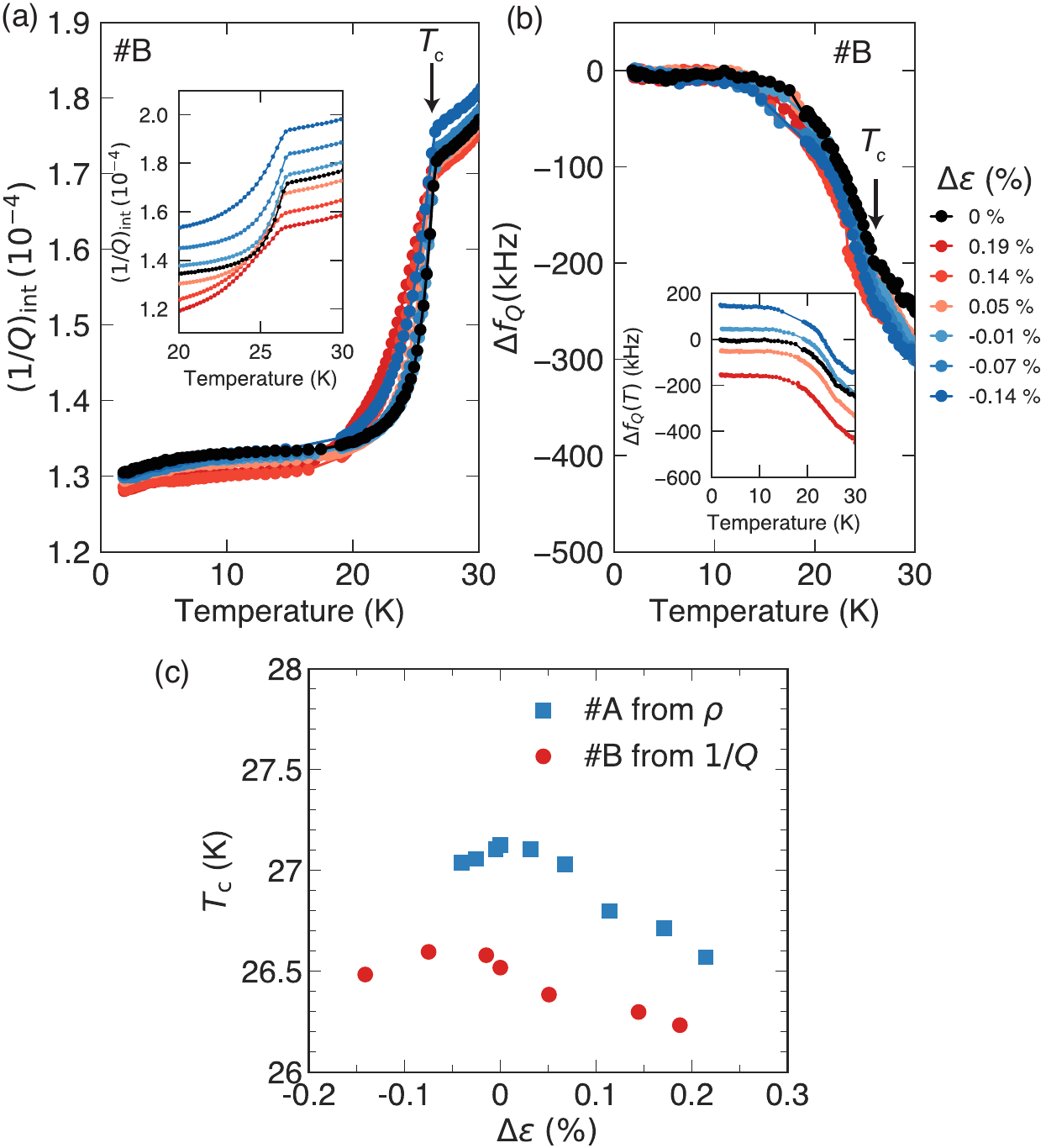}% Here is how to import EPS art
    \caption{\label{fig:MW}  (a) and (b) Temperature dependence of the microwave resonance parameters of BaFe$_2($As$_{1-x}$P$_x)$$_2$ (sample $\#$B, $x \sim 0.38$) under strain: $1/Q_{\rm int}$ (a) and $\Delta f_{Q} (T)$ (b). Black arrows indicate the superconducting transition. Insets show the same data with vertical offset for clarity. (c) Strain dependence of the superconducting transition temperature, determined by both electrical resistivity (blue squares) and microwave loss (red circles).}
  \end{figure}

  To track the microwave properties of the sample, we measured the resonance of the rutile cavity in the presence of the sample. Figure \ref{fig:MW}(a) shows the temperature dependence of $Q_{\rm int}$. $Q_{\rm int}$ consists of several factors in addition to the sample: $1/Q_{\rm int} = 1/Q_{\rm s}+1/Q_{\rm BG}$, where $ 1/Q_{\rm s}$ and $1/Q_{\rm BG}$ represent the loss due to the sample and other backgrounds, respectively (see also supplementary material).
  % $1/Q_{\rm BG}$ includes the radiation loss, the resistivity of metals other than sample, the dielectric loss of rutile and quartz.
  The complicated structure of the current strain-tunable cavity makes it difficult to quantitatively assess $1/Q_{\rm BG}$ for subtraction. However, the sharp reduction of $1/Q_{\rm int}$ around 27 K clearly indicates the superconducting transition, dominated by the sample signal $ 1/Q_{\rm s}$, as shown in Fig. \ref{fig:MW}(a). The signature of the superconducting transition can be also observed as a kink in the resonance frequency shift $\Delta f_Q (T)$ from the minimum measured temperature, with the background contribution subtracted, as shown in Fig. \ref{fig:MW}(b). This clear identification of the superconducting transition enables us to determine the strain dependence of $T_{\rm c}$ through these two microwave parameters.

    The induced strain in our strain-tunable cavity setup is evaluated based on the displacement difference $\Delta L_{\rm piezo} = \Delta L_{\rm inner} - \Delta L_{\rm outer}$ between the inner and outer piezo actuators, using the resistive strain gauges attached to the surface of each piezo actuator. Induced strain in the platform (approximately equal to that in the sample) is essentially calculated by dividing $\Delta L_{\rm piezo}$ by the effective length of the platform $l_{\rm eff}$ that represents the characteristic length scale where the strain is concentrated\cite{10.1063/5.0008829}:$\Delta \varepsilon = \Delta L_{\rm piezo} /l_{\rm eff}$. In practice, the induced strain, especially in the large strain region, exhibits nonlinear and hysteresis behavior with respect to the applied displacement of the piezo actuator\cite{10.1063/5.0008829}. Therefore, we experimentally calibrated the relationship between $\Delta L_{\rm piezo}$ and the actual strain $\Delta \varepsilon$ of the quartz platform in advance by directly attaching a resistive strain gauge, later removed to avoid interference with the microwave resonance measurements of the sample. 
    % It should be noted that, for the conventional resistivity measurements shown in Fig. \ref{fig:BaFe2As2}(b), the strain was monitored in situ using this strain gauge glued to the backside of the platform.

  % The strain dependence of these two microwave parameters reveal that these superconducti n transition temperatures are suppressed by both compressive and tensile strain, which is consistent with our resistivty measurements.
  
  % \begin{figure}
  %   \includegraphics[width=0.48\textwidth]{Fig4.pdf}% Here is how to import EPS art
  %   \caption{\label{fig:Tc} Description of strain-microwave cavity.}
  % \end{figure}

  Suppression of $T_{\rm c}$ under both compressive and tensile strain is consistently observed in both resistivity and microwave resonance measurements, as shown in Fig. \ref{fig:MW}(c). This result demonstrates that our strain-tunable microwave cavity successfully probes the strain-controlled superconducting state, highlighting its potential as an effective tool for exploring emergent physical phenomena. Although we focused on strain-tuned superconducting phase diagram within the current scope of this study, the strain-dependent microwave response contains further information on the sample properties. For instance, in the superconducting state, since shifts in resonance frequency $\Delta f_{\rm Q} (T)$ contain essential information about superconducting quasiparticle excitations. More precise measurements could elucidate strain-induced changes in superconducting gap structure, offering a promising direction for future studies on potential multiple superconducting phases such as those in UPt$_3$\cite{RevModPhys.74.235} and Sr$_2$RuO$_4$\cite{Grinenko2021}.

  % For instance, the loss $1/Q_{\rm total}$ above $T_{\rm c}$ increases under compressive (negative) strain, which is consistent with trend observed in the DC resistivity behavior. Strain response of resistivity is dominated by the strain-induced changes in resistivity anisotropy, indicating our $1/Q_{\rm total}$ probes anisotropy of surface impedance $R_x - R_y$, where $R_{i = x,y,z}$ represents the surface resistance with current running along the $i$ direction. This finding is surprising because, in principle, it is difficult to extract the anisotropy of surface resistance from our measured loss, because it is the sum, not the diffrence, of the inplane surface resistance\cite{PhysRevLett.73.2484}. The measured sample is a thin rectangle: strain is applied along the longer axis ($x$), the shorter axis is $y$, and the thickness is $z$. This elongated shape of our sample ($L_x/L_y \sim 2.6$) results in a higher dominance of $ R_{x}$ contribution as $J_{\rm s} L_z (L_x R_{x} + L_y R_{y})$\cite{PhysRevLett.73.2484}, leading to behavior similar to that observed in DC elastoresistance. Here, $J_s$ is the surface current density, and $L_x$,$L_y$,$L_z$ are dimensions of the sample. 

  \begin{figure}
    \includegraphics[width=0.4\textwidth]{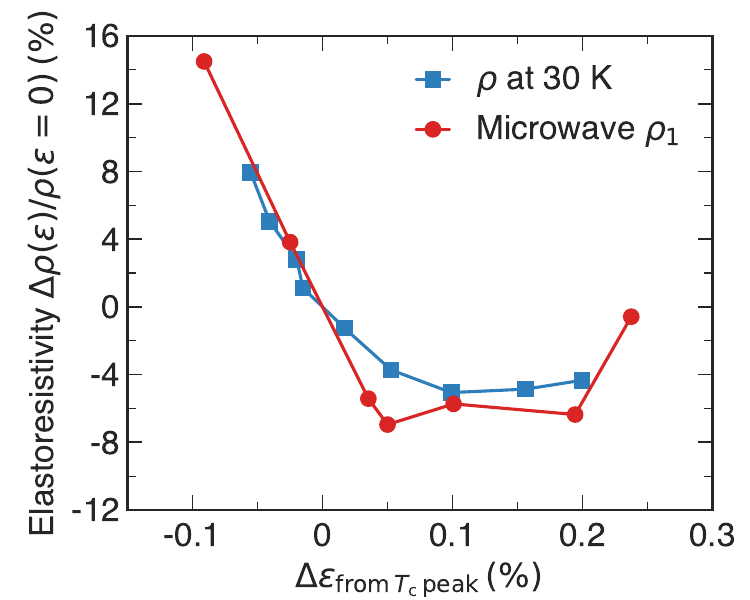}% Here is how to import EPS art
    \caption{\label{fig:Elasto} Comparison of standard and microwave elastoresistivity just above $T_{\rm c}$: Red circles and blue squares represent strain-induced changes in microwave resistivity $\rho_1$ and DC-limit resistivity, respectively. The zero-point strain is defined as the strain value at the peak of $T_{\rm c}$ to eliminate the thermal strain offsets observed across different measurement runs.}
  \end{figure}

Beyond the superconducting regime, strain-dependent $R_{\rm s}$ in the normal state can offer further insights into the electronic anisotropy of the material. This is conceptually analogous to elastoresistivity observed in standard transport, which has been widely used, particularly as a measure of nematic susceptibility in iron-based superconductors\cite{doi:10.1126/science.1221713,doi:10.1126/science.aab0103,doi:10.1073/pnas.1605806113,doi:10.1073/pnas.2110501119}. Although the complicated contributions of the background loss make it challenging to isolate the loss from $R_s$ of the sample, we successfully extracted the contribution from $R_s$ using the following procedure. The residual $R_{\rm s}$ is negligibly small \cite{doi:10.1126/science.1219821} ($R_{\rm s}$(4.2K)/$(R_s (T_{\rm c})$ at 4.9\,GHz is less than 0.1\% for $x = 0.3$), and thus $\Delta (1/Q_{\rm int})_{\rm SC} = (1/Q_{\rm int})_{T=T_{\rm c}} -(1/Q_{\rm int})_{T=T_{\rm min}} $ can be attributed purely to $R_{\rm s}$. By analyzing the strain dependence of  $\Delta (1/Q_{\rm int})_{\rm SC}$, we can evaluate the strain-induced changes in $\rho_1 = 2 R_{\rm s}^2 / \mu_0 \omega$, which we define as microwave surface elastoresistivity, just above $T_{\rm c}$, as shown in Fig. \ref{fig:Elasto}. To validate the successful observation of microwave elastoresistivity, we directly compared it with the standard elastoresistivity measurements based on the DC-limit resistivity $\rho$ at 30 K, and found excellent agreements between the two methods. Both measurements reveal a non-monotonic evolution of elastoresistivity, similar to that observed in Co-doped BaFe$_2$As$_2$\cite{PaulNatPhys}. For comparison, we re-centered the zero strain point at the peak of $T_{c}$ based on Fig. \ref{fig:MW}(c), thereby eliminating offset thermal strain effects observed across different runs. In the small strain region, the elastoresistivity exhibits a steep negative slope ($\frac{{\rm d} \Delta \rho_1(\varepsilon)/\rho_1(\varepsilon=0)}{{\rm d} \varepsilon} \sim -155$), consistent in both sign and magnitude with the nematic susceptibility expected slightly away from the nematic quantum critical point\cite{doi:10.1126/science.aab0103}. These results support the idea that microwave elastoresistivity provides a useful, contactless probe of the electronic anisotropic properties.

  \begin{figure}
    \includegraphics[width=0.48\textwidth]{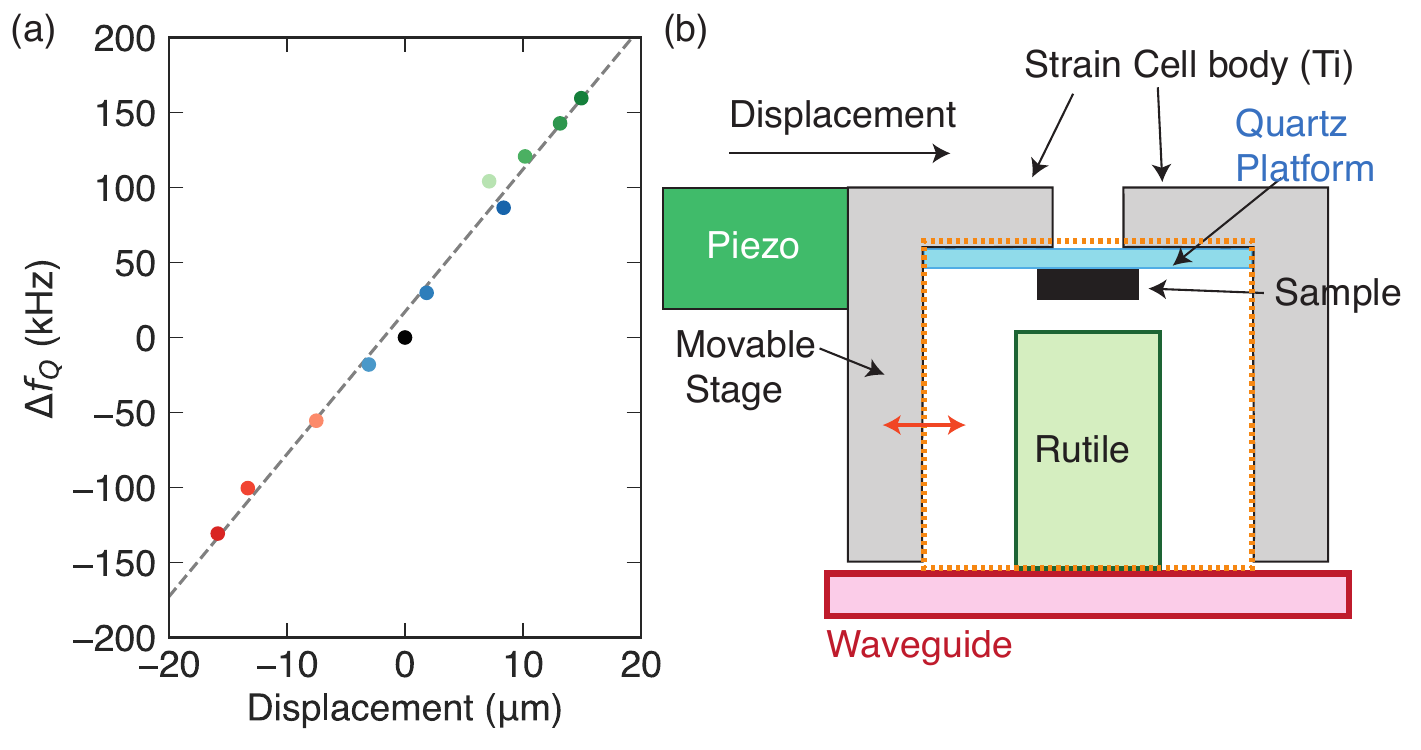}% Here is how to import EPS art
    \caption{\label{fig:Freq} Effects of strain cell movement on resonance frequency. (a) Resonance frequency shift shows a linear dependence on the displacement applied to the strain cell. (b) Schematic representation of the enclosure environment of the rutile cavity. The area enclosed by the dashed orange lines highlights the effective volume of the cavity. The movement of the stage alters this enclosure volume, causing the resonance frequency shift: the displacement in positive (negative) direction reduces (increases) the enclosure volume and, as a result, increases (decreases) the resonance frequency.}
  \end{figure}

An additional functionality of our developed strain-tunable microwave cavity is its sensitivity to mechanical displacement. We found a linear relationship between resonance frequency and the displacement applied by piezo actuators, as shown in Fig. \ref{fig:Freq}(a). This frequency shift arises from subtle changes in enclosure environment around the rutile due to the movement of the strain cell body. Notably, the magnitude of this frequency shift is comparable to that induced by temperature changes associated with the superconducting transition of the sample (see Fig. \ref{fig:MW}(b)), which makes it particularly challenging to extract the strain dependence of $X_{\rm s}$. Conversely, this pronounced frequency shift can be leveraged as a sensitive probe of the strain-cell movement itself. By calibrating the linear relationship between frequency and displacement, the applied strain can be inferred directly from the microwave signal. This capability suggests the potential for an all-microwave readout system, eliminating the need for additional resistive strain gauges used in this study.

   Finally, we discuss potential applications of our mechanically tunable microwave cavity. This tool enables in situ control and monitoring of superconducting pairing states by applying uniaxial stress and measuring microwave surface impedance of superconductors. Furthermore, this system offers a method to explore deeper regions of phase diagram of superconductors, including potential strain-induced coexisting order within the superconducting phase. This is particularly relevant for many unconventional superconductors, which often emerge in the vicinity of other ordered phases such as magnetism. In addition, the robustness of the rutile cavity against high magnetic fields\cite{10.1063/1.2167127} expands its applicability to  exploring wide regions of  electronic phase diagrams. Applications to spin systems represent a promising avenue for further development. For instance, the utility of the magnetoelastic effect works for controlling the magnetization direction, and such changes can be probed using ferromagnetic magnetic resonance. Overall, our strain-tunable microwave cavity serves as a versatile platform for investigating potential applications in various quantum hybrid systems\cite{Lachance-Quirion_2019} as well as studying strain-sensitive phenomena across a wide range of quantum materials.
   
   In summary, we developed the microwave cavity hybridized with the piezoelectric-based strain device. Using this system, we measured strain-dependent microwave surface impedance through changes in the quality factor and resonance frequency shifts in iron-based superconductor BaFe$_2($As$_{1-x}$P$_x)$$_2$, demonstrating the observation of strain-controlled superconductivity. A comparative study using standard resistivity measurements confirms that our microwave spectroscopy measurements successfully pinpoint the intrinsic properties of superconductors. In particular, we extracted the strain-induced changes in $R_{\rm s}$ of the superconductor, extending the concept of the elastoresistivity to the microwave regime. The observed microwave elastoresistivity exhibits strain dependent behavior consistent with the conventional elastoresistivity, offering an alternative contactless method to investigate anisotropic properties of materials. By combining mechanical tuning and microwave resonance spectroscopy measurements, this system paves the way for studying exotic superconducting states and serves as a promising platform for hybrid quantum functional systems. 
   
\vspace{3mm}
\noindent\textbf{Supplementary Material}\\
Further details on the measurement and analysis of rutile resonance are provided in the supplementary material.

\begin{acknowledgments}
 We thank M. Roppongi with helpful discussions. We also thank M. Yashima for providing the cuprate superconductors used during the development of the strain-tunable cavity. This work was supported by Grants-in-Aid for Scientific Research (KAKENHI) (Grans No. JP22K18690, JP23K17879, and JP23K23207), Toyota Riken Scholar Program, and the Multidisciplinary Research Laboratory System (MRL), Osaka University.
\end{acknowledgments}

\section*{AUTHOR DECLARATIONS}
\subsection*{Conflict of Interest}
The authors have no conflicts to disclose.

\subsection*{Author Contributions}
\textbf{Suguru Hosoi:} Conceptualization (equal); Data curation (lead); Formal analysis (lead); Funding acquisition (lead); Investigation (lead); Methodology (lead); Resources (lead); Writing - original draft - (lead); Writing - review \& editing - (equal).
\textbf{Kohei Matsuura:} Conceptualization (equal); Formal analysis (support); Funding acquisition (support); Investigation (support); Methodology (support); Resources (support); Writing - review \& editing - (equal).
\textbf{Masaaki Shimozawa:} Funding acquisition (support); Investigation (support); Methodology (support); Resources (support); Writing - review \& editing - (equal).
\textbf{Koichi Izawa:} Funding acquisition (support); Resources (support); Writing - review \& editing - (support).
\textbf{Shigeru Kasahara:} Resources (support); Writing - review \& editing - (support).
 \textbf{Takasada Shibauchi:} Formal analysis (support); Resources (support); Writing - review \& editing - (support).

\section*{DATA AVAILABILITY}
The data that support the findings of this study are available from the corresponding authors upon reasonable request.

\nocite{*}
\bibliography{aipsamp}% Produces the bibliography via BibTeX.

\end{document}